\documentclass[twocolumn,showpacs,preprintnumbers,amsmath,amssymb]{revtex4}
\usepackage{graphicx}
\begin{document}

\title{Phase diagram of ultracold atoms in
optical lattices: Comparative study of slave fermion and slave
boson approaches to the Bose-Hubbard model}

\author{Yue Yu $^{1,2}$ and S. T. Chui$^2$}
\affiliation{1. Institute of Theoretical Physics, Chinese Academy
of Sciences, P.O. Box 2735, Beijing 100080, China} \affiliation{2.
Bartol Research Institute, University of Delaware, Newark, DE
19716}
\date{\today}
\begin{abstract}
We perform a comparative study of the finite temperature behavior
of ultracold Bose atoms in optical lattices by the slave
fermion and the slave boson approaches to the Bose Hubbard model. The
phase diagram of the system is presented.
Although both approaches are equivalent without
approximations, the mean field theory based on the slave fermion
technique is quantitatively more appropriate. Conceptually, the
slave fermion approach automatically excludes the double occupancy
of two identical fermions on the same lattice site.
By comparing to known results in limiting cases, we find the
slave fermion approach better than the slave boson
approach.  For example, in the non-interacting limit, the critical
temperature of the superfluid-normal liquid transition calculated
by the slave fermion approach is closer
to the well-known ideal Bose gas result.
At zero-temperature limit of the critical interaction strength
from the slave fermion approach is also closer
to that from the direct calculation using a
zero-temperature mean field theory.
\end{abstract}

\pacs{03.75.Lm,67.40.-w,39.25.+k}

\maketitle

\section{Introduction}

Strongly correlated systems are of longstanding interest in
studies of condensed matter physics. Ultra-cold
atoms in optical lattices offer new opportunities to study
strongly correlated phenomena in a highly controllable
environment\cite{1,2,3,kohl}. A quantum phase transition,
the superfluid/Mott-insulator transition, was demonstrated using $^{87}$Rb
atoms in three- \cite{2} and one-dimensional lattices \cite{kohl}.
Strongly correlated phenomena for boson systems may be studied
theoretically by the Bose-Hubbard model \cite{BH}.
Experimental feasibility was established by microscopic
calculations of the model parameters for cold boson atoms in
optical lattices \cite{jaks}. A review of recent works on the
superfluid-insulator quantum phase transition at zero temperature
is given in ref. \cite{zw}.

Strictly speaking, a quantum phase transition can not be observed
at any finite temperature. The experimental data give only a
signal that the system is towards a quantum phase transition if
the temperature is extrapolated to zero. What the experiments
really observed was a transition from the superfluid to the normal
liquid whose compressibility is very close to zero and the
system is  practically a Mott insulator. Such a 'classical' phase
transition has been investigated recently by Dickerscheid et al
\cite{dick}. Phase diagrams for a given atom density were
calculated in the temperature-interaction plane and the chemical
potential-interaction plane: For a commensurate optical lattice,
there are only the superfluid and the Mott insulator phases at zero
temperature. At finite temperatures, starting from the superfluid
phase, there is a superfluid/normal liquid phase transition
while the Mott insulator phase crossovers to the normal liquid.

In order to extend the ordinary mean field approach for the Bose
Hubbard model \cite{mean} to include the finite temperature
effects, the slave boson technique \cite{kot} was used. The slave
particle technique has been widely applied in dealing with the
strongly correlated electron systems \cite{sbsf}. In principle,
the slave boson and slave fermion approaches are equivalent.
However, in practical calculations, approximations still have to
be used. It was well-known that in the $t$-$J$ model, the same
mean field approximation using the slave boson or slave fermion
leads to very different phase diagrams \cite{sbsf}. It was known
that the slave boson mean field approximation can qualitatively
describe the phase diagram of the cuprates at finite doping.
However, due to the Bose condensation of the holons, the slave
boson mean field approximation does not produce correctly the
ferromagnetic Mott insulator phase. One of the purpose of this
work is to examine if both slave particle approaches to the Bose
Hubbard model give the same physical results under the mean field
approximation.

In these approaches, there is a constraint that each site can be
occupied by only one slave particle. With this exact constraint,
the model is very hard to solve. A standard approximation is to
relax the constraint on each site to the requirement that the
average slave particle per site over the lattice be equal to 1.
While both approaches give the same qualitative phase diagram, we
shall see that the quantitative behaviors derived from the slave
fermion approach are more accurate. The advantage of the slave
fermions is that the Fermi statistics automatically excludes two
same type slave fermions from occupying the same site even when
the constraint is relaxed. We shall see that the configurations
with the multi-occupations of different types of slave fermion are
far away from the mean field state we consider. Thus, these
configurations will not significantly influence our results.
However, the statistics of the slave particle affect the result
remarkably. For repulsive interactions, there are two
unsatisfactory features arising from the finite temperature mean
field theory in the slave boson approache\cite{dick}. One of them
is that the critical on-site repulsive $U_c\approx 5.83$ from a
direct zero temperature mean field calculation and differs  from
the critical $U_c\approx 6$ in $T\to 0$ from the finite
temperature mean field. This difference is much smaller with the
slave fermion approach. The other was that there is a maximum
$T_c$ at $U\ne 0$ in the $U$-$T_c$ curve which is obviously
unphysical.  We find that these deficiencies are corrected in the
slave fermion approach. Furthermore, for $U=0$, the critical
temperature of the superfluid-normal liquid phase transition for
the ideal Bose gas was well-known. We find that this critical
temperature calculated by the slave fermion approach is much
closer to its exact value than that by the slave boson approach.

This paper is organized as follows. In Sec. II, we give an
overview of our slave particle approaches. In Sec. III, the
perturbation theory is introduced. In Sec. IV, we give our main
results according to the mean field theory. Section V is our
conclusion.

\section{Slave particle approach}

 A boson operator
on site $i$ may be expressed by the occupation state
$|\alpha\rangle$, i.e.,
\begin{eqnarray}
a_i^\dagger=\sum_\alpha\sqrt{\alpha+1}|\alpha+1\rangle_{ii}\langle
\alpha|.
\end{eqnarray}

The slave particles are the auxiliary particles, which are
obtained by mapping the occupation state $|\alpha\rangle_i\to
a_{\alpha,i}$. According to this mapping, the original boson
creating operator $a^\dagger_i$ on the lattice site $i$ may be
decomposed as
\begin{eqnarray}
a^\dagger_i=\sum_{\alpha=0}\sqrt{\alpha+1} a_{\alpha+1, i}^\dagger
a_{\alpha,i},
\end{eqnarray}
where $a_{\alpha,i}$ may be either the (slave) boson operator
$b_{\alpha,i}$ with
$[b_{\alpha,i},b^\dagger_{\beta,j}]=\delta_{\alpha\beta}\delta_{ij}$
or the (slave) fermion operator $c_{\alpha,i}$ with
$\{c_{\alpha,i},c^\dagger_{\beta,j}\}=\delta_{\alpha\beta}\delta_{ij}$.
As the auxiliary particles, they have to obey the constraint
\begin{eqnarray}
\sum_\alpha n^\alpha_i=\sum_\alpha
a^\dagger_{\alpha,i}a_{\alpha,i}=1, \label{cst}
\end{eqnarray}
on each site, which corresponds  to the completeness of the states
:$\sum_\alpha |\alpha\rangle\langle\alpha|=1$ and the original
Bose commutation relation: $[a_i,a_j^\dagger]=\delta_{ij}$.

The Bose Hubbard Hamiltonian we will focus on reads
\begin{eqnarray}
H=-t\sum_{\langle ij\rangle}a^\dagger_ia_j-\mu\sum
n_i+\frac{U}2\sum_i n_i(n_i-1),
\end{eqnarray}
where the symbol $\langle ij\rangle$ denotes the sum over all
nearest neighbor sites. $\mu$ is the chemical potential. The
hopping amplitude $t$ and the on-site interaction $U$ are defined
by
\begin{eqnarray}
t&=&\int d {\bf r} W^*({\bf r})[-\frac{\hbar^2}{2m}\nabla^2+V({\bf
r})] W({\bf r}+{\bf a}),\nonumber\\
U&=&g\int d{\bf r}|W({\bf r})|^4,
\end{eqnarray}
where $g=\frac{4{\pi}a_s\hbar^2}{m}$ with $a_s$ the $s$-wave
scattering length of the atoms and $m$ the mass of atom; $W({\bf
r})$ is the Wannier function corresponding to the lowest Bloch
band and $V({\bf r})$ is the periodic optical lattice potential.
In three dimensions, taking the periodic potential with a form
\begin{eqnarray}
V({\bf r})=V_0(\sin^2(kx)+\sin^2(ky)+\sin^2(kz)),
\end{eqnarray}
we can calculate $U$ and $t$ from the band theory \cite{jaks}. In
Fig. 1, we display the relation between $U/6t$ and $V_0$.

\begin{figure}
\begin{center}
\includegraphics[width=8cm]{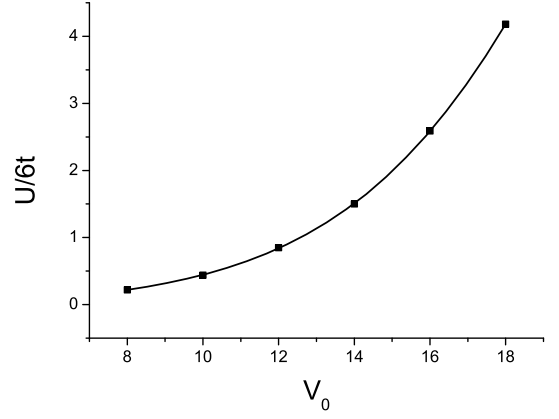}
\end{center}
 \caption{\label{fig:Fig. 1}
 The dependence of the Bose Hubbard parameter $a_s U/6t$ for the three
 dimensional optical lattice as a function of the strength of
 the lattice potential $V_0$. The scattering length $a_s$ is in units of a
 nanometer. $V_0$ is in units of the recoil energy
$E_r=\hbar^2 k^2/2m$}
\end{figure}

In the slave particle language,
the one-component Bose Hubbard Hamiltonian in a $d$-dimensional
cubic lattice with $L$ sites reads
\begin{eqnarray}
H&=&-t\sum_{\langle
ij\rangle}\sum_{\alpha,\beta}\sqrt{\alpha+1}\sqrt{\beta+1}
a^\dagger_{\alpha+1,i}a_{\alpha,i}a^\dagger_{\beta,j}
a_{\beta+1,j}\nonumber\\
&-&\mu\sum_i\sum_\alpha\alpha
n^\alpha_i+\frac{U}{2}\sum_i\sum_\alpha\alpha(\alpha-1)n_i^\alpha,
\end{eqnarray}
With the constraint (\ref{cst}), the standard path integral leads
to the partition function of the system
\begin{eqnarray}
&&Z=Tre^{-\beta H}=\int Da_\alpha D\bar a_\alpha D\lambda ~e^{-S_E},\nonumber\\
&&S_E[\bar a_\alpha, a_\alpha,\lambda]=\int_0^{1/T} d\tau
\biggl\{\sum_i\sum_{\alpha} \bar
a_{\alpha,i}[\partial_\tau-\alpha\mu\nonumber\\
&&~~+\frac{U}2\alpha(\alpha-1)
-i\lambda_i]a_{\alpha,i}+i\sum_i\lambda_i\\
&&~~-t\sum_{\langle
ij\rangle}\sum_{\alpha\beta}\sqrt{\alpha+1}\sqrt{\beta+1} \bar
a_{\alpha+1,i}a_{\alpha,i}\bar a_{\beta,j}
a_{\beta+1,j}\biggr\},\nonumber
\end{eqnarray}
where we have taken $\hbar=k_B=1$. $\bar a_{\alpha i}$ is equal to
the complex conjugate $b^*_{\alpha,i}$ of the bosonic field
$b_{\alpha,i}$ for the slave boson while being the Grassmann
conjugate $\bar c_{\alpha i}$ of the fermion field $c_{\alpha i}$
for the slave fermion. The integrals over the Lagrange multiplier
field $\lambda_i(\tau)$ come from the constraint (\ref{cst}):
\begin{eqnarray}
&&\prod_i\delta(\sum_\alpha n^\alpha_i-1)=\int D\lambda
\exp\biggl[i\int_0^\beta\sum_i\lambda_i\nonumber\\
&&\times(\sum_\alpha n^\alpha_i-1)d\tau\biggr].
\end{eqnarray}
In the sense of the $\delta$-function, the $\lambda_i$ fields have to
be real to ensure the constraint is correctly taken into account.

To decouple the four slave particle term in the Hamiltonian, we
introduce a Hubbard-Stratonovich field $\Phi_i$ which is a bosonic
field and may be identified as the order parameter of the
superfluid. The integral
\begin{eqnarray}
&&\int D\Phi D\Phi^* \exp\biggl[-\int d\tau
t\sum_{\langle ij\rangle}\nonumber\\
&&(\Phi^*_i- \sum_\alpha\sqrt{\alpha+1}\bar
a_{\alpha+1,i}a_{\alpha,i})\nonumber\\
&&\cdot(\Phi_j- \sum_\alpha\sqrt{\alpha+1}\bar
a_{\alpha,j}a_{\alpha+1,j})\biggr],
\end{eqnarray}
is obviously a constant. The partition function can
be written as
\begin{eqnarray}
&&Z=\int D\Phi D\Phi^* D\bar a_\alpha D a_\alpha D\lambda
e^{-S_{eff}[\Phi,a_\alpha,\lambda]},\nonumber\\
&&S_{eff}[\Phi,a_\alpha,\lambda]=\int d\tau
\biggl[\sum_i\sum_{\alpha} \bar
a_{\alpha,i}[\partial_\tau-\alpha\mu\nonumber\\
&&~~+\frac{U}2\alpha(\alpha-1)
-i\lambda_i]a_{\alpha,i}+i\sum_i\lambda_i\\
&&~~+t\sum_{\langle
ij\rangle}(\Phi_i^*\Phi_j-\Phi_i^*\sum_\alpha\sqrt{\alpha+1}\bar
a_{\alpha,j}a_{\alpha+1,j}\nonumber\\
&&~~-\Phi_j\sum_\alpha\sqrt{\alpha+1}\bar
a_{\alpha+1,i}a_{\alpha,i})\biggr]. \label{seff}
\end{eqnarray}
This effective action is the starting point of the slave particle
approach. The slave boson and the slave fermion representations are
only reexpression of the original Bose Hubbard model. Both of them
are equivalent to the original model before any approximation. So
far, all formal transformations we made are rigorous.

\section{perturbation theory}

The physical meaning of the $\Phi$ field may be seen by using its
equation of motion:
\begin{eqnarray}
\langle\Phi_i\rangle=\langle\sum_\alpha\sqrt{\alpha+1}\bar
a_{\alpha,i}a_{\alpha+1,i}\rangle=\langle a_i\rangle.
\end{eqnarray}
This means that $\Phi_i$ indeed serves as an order parameter
field. Near the Mott transition, this order parameter is small and
one can use perturbation theory to solve the system described by
the action (\ref{seff}). The difficulty is that there is no
way to exactly solve the problem if the $\lambda$ field varies from
site to site.  A widely-used
approximation is to relax the constraint (3) by replacing the local
constraint Lagrange multiplier $\lambda_i(\tau)$ by an imaginary
time- and site-independent field $\lambda$. That is, relaxing
the condition of exactly  one slave particle per site to
one with an average of one particle per site. It
implies that multi-occupation of the slave particles on the same
site is allowed.  For slave bosons, this relaxation allows the
same type of the boson to multi-occupy a single site. However, for
the same type of slave fermions,  multi-occupation of the same
site is automatically forbidden by the Pauli principle. The value
of $\lambda$ will be variationally determined.

To do the perturbation calculation, it is convenient to make a
Fourier transformation for the fields $A_i=a_i, \Phi_i$ and
$\lambda_i$:
\begin{eqnarray}
&&A_i=\frac{1}{\sqrt{L\beta}}\sum_{k,n}A_{\alpha,kn}e^{i{\bf
k}\cdot {\bf i}-i\omega_n\tau},
\end{eqnarray}
where the Matsubara frequencies $\omega_n=2\pi n T$ for bosonic
fields and $\omega_n=(2n+1)\pi T$ for fermionic fields. The
approximation of the site-independent of $\lambda_i$ implies that
all $\lambda_{k,n}=\lambda_{0,0}=\lambda\sqrt{L\beta}$.

After all these preparations and then some algebra,
we arrive the effective action to second order of the order parameter field,
\begin{eqnarray}
S_{E,{\rm eff}}[\Phi^*,\Phi]=\beta\Omega_0-\sum_{{\bf
k},n}\Phi^*_{{\bf k},n}G^{-1}({\bf k},i\omega_n)\Phi_{{\bf k},n},
\end{eqnarray}
where $\Phi_{{\bf k},n}$ is the Fourier component of $\Phi_i$ and
the zero order thermodynamic potential is given by
\begin{eqnarray}
\Omega_0=iL\lambda\mp\frac{L}\beta\sum_\alpha \ln(1\mp e^{-\beta
\epsilon_0(\alpha)}),
\end{eqnarray}
 where $\beta=1/T$ and
$\epsilon_0(\alpha)=-i\lambda-\alpha\mu+\alpha(\alpha-1)U/2$ and
$-$ ($+$) sign corresponds to the slave boson ( slave fermion)
approximation. The Green's function is defined by
\begin{eqnarray}
-G^{-1}({\bf k},i\omega_n)=\epsilon_{\bf k}+\epsilon_{\bf
k}^2\sum_\alpha(\alpha+1)\frac{n^\alpha-n^{\alpha+1}}{i\omega_n+\mu
-\alpha U}, \label{green}
\end{eqnarray}
where the dispersion $\epsilon_{\bf k}=2t\sum_i\cos(k_ia)$ with
$z$ the nearest neighbor partition and ${\bf a}$ the lattice
spacing vector of a $d$-dimensional cubic lattice. The bosonic
Matsubara frequency $\omega_n=\frac{2\pi n}\beta$. The slave
particle occupation number is given by
\begin{eqnarray}
n^\alpha=\frac{1}{\exp\{\beta[-i\lambda-\alpha\mu+\alpha(\alpha-1)U/2]\}\mp
1}, \label{oc}
\end{eqnarray}
corresponding to the slave boson and slave fermion, respectively.

\section{mean field theory}

We focus on repulsive interactionis with $U>0$ in this paper.
According to Landau theory, the condition $G^{-1}({\bf 0},0)=0$
may be used to determine the critical point of the phase
transition between the superfluid and the normal liquid
\cite{dick}.

The key difference between the slave boson and slave fermion is
their quantum statistics, which leads to the sign difference $\mp$
in equation (18). The difference appears because we have
approximated all $\lambda_i(\tau)$ by a real constant
$\lambda$. We discuss the zero temperature and the finite temperature
cases separately.

\subsection{Zero Temperature}

In the zero temperature limit, Dickersheid et al \cite{dick}
investigated commensurate fillings and  assume the number
of particles of each well to be fixed at some value $\alpha'$:
$n^{\alpha'}=\delta_{\alpha,\alpha'}$ in mean field theory. This
mean field assumption works for both kinds of slave particles. In
terms of $G^{-1}({\bf 0},0)=0$ and the Green's function
(\ref{green}), it is easy to calculate the phase boundaries in the
$\mu$-$U$ plane \cite{dick}: The Mott insulator phase is in the
regimes where $\bar\mu$ lies between $\bar \mu^{\alpha'}_\pm$
\begin{eqnarray}
\bar\mu^{\alpha'}_\pm=\frac{1}2[\bar U(2\alpha'-1)-1]\pm\frac{1}2
\sqrt{\bar U^2-2\bar U(2\alpha'+1)+1}. \label{ml}
\end{eqnarray}
Here $\bar\mu=\mu/zt$ and $\bar U=U/zt$. It is easy to check that
for $U>0$, $\mu$ is positive. Eq. (\ref{ml}) reproduce results of
previous mean field studies \cite{mean}. For the first Mott lobe,
$\alpha'=1$, the critical temperature is give by the zero of the
square root in (\ref{ml}), which gives $\bar U_c\approx 5.83$. The
quantum Monte Carlo calculation showed the critical $U_c$ is
somewhat smaller: $\bar U_c\approx 4$ \cite{kz}.

\begin{figure}
\begin{center}
\includegraphics[width=7.5cm]{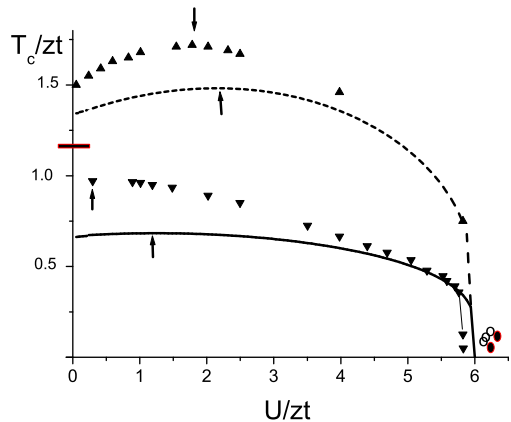}
\end{center}
 \caption{\label{fig:Fig. 2}
 The critical temperature of the normal-superfluid
 phase transition as a function of the interaction $U$.
 The solid curve is for the $\alpha=0,1,2$ slave fermions.
 The dash curve is for the $\alpha=0,1,2$ slave bosons.
 The down-triangles for the $\alpha=0,1,2,3$ slave fermions.
 The up-triangles for the $\alpha=0,1,2,3$ slave bosons.
 The data for the slave bosons are taken from \cite{dick}.
 The arrows indicate the position of the maxima of the critical
 temperature.
 The longer bar on the $\bar T_c$-axis is the critical temperature
 ($\bar T_c^{ideal}=1.18$) for the ideal Bose gas in three-dimensional lattice. }
\end{figure}

\begin{figure}
\begin{center}
\includegraphics[width=7.5cm]{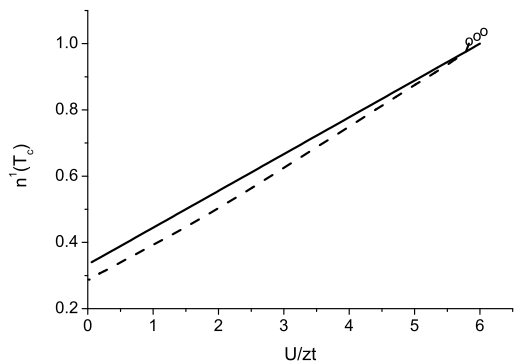}
\end{center}
 \caption{\label{fig:Fig. 3} The $\alpha=1$ slave
 fermion number versus $U/zt$ at the critical point of the phase
 transition. The solid curve is for three slave fermions and the dash line is for
  four slave fermions. The empty circles represent the unphysical
   results with $n^1>1$ at the "critical temperature". }
\end{figure}

\subsection{Finite Temperature}

We next turn our attention to the finite temperature behavior. For
an site-independent parameter $\lambda$, the relaxed constraint
associated with (\ref{cst}) may be derived by taking a saddle
point approximation for $\lambda$, i.e., minimizing the
thermodynamic potential: $\partial \Omega/\partial \lambda=0$. The
particle conservation condition reads $-\partial \Omega/\partial
\mu=N$. The corresponding saddle point equations are given by
\begin{eqnarray}
&&N_s\sum_\alpha (1- n^\alpha)-\frac{i}\beta\sum_{k,n}G({\bf
k},i\omega_n) \frac{\partial
G^{-1}({\bf k},i\omega_n)}{\partial\lambda}=0,\nonumber\\
&&N_s\sum_\alpha\alpha n^\alpha+\frac{1}\beta\sum_{k,n}G({\bf
k},i\omega_n) \frac{\partial G^{-1}({\bf
k},i\omega_n)}{\partial\mu}=N.
\end{eqnarray}
In these equations, all the possible multi-occupant states are
included.

The mean field approximation implies the last terms in the above
two equations may be neglected, which gives
\begin{eqnarray}
\sum_{\alpha=0} n^\alpha=1, ~~{\rm and}~~\sum_\alpha\alpha
 n^\alpha=N/L.\label{cstde}
\end{eqnarray}
From the first equation, we see that all multi-occupancy of the
different types slave particles are excluded in the mean field
approximation \cite{note1}. The difference between the slave boson
and slave fermion approaches is shown only on their different
statistics. The critical point of the superfluid/normal liquid
phase transition, in terms of $G^{-1}({\bf 0},0)=0$, is determined
by
\begin{eqnarray}
\sum_\alpha (\alpha+1)\frac{n^{\alpha+1}-n^\alpha}{\bar\mu-\alpha
\bar U}=1. \label{critical}
\end{eqnarray}
We now restrict to the commensurate state and take the total
density to $n=N/L=1$. To be able ito solve eqs. (\ref{cstde}) and
(\ref{critical}), one has to make a cut-off in $\alpha$. For a cut
off $\alpha_M=2$ so that only $\alpha=0,1,2$ are allowed, these
equations can be analytically solved. The resulting equation have
been solved by Dickerscheid et al in the slave boson case
\cite{dick}. They are also easily solvable for the slave
fermion case. For a given $\bar U$, $\mu=\bar U/2$
and $n^1=(\bar U+3)/9$ for both kinds of slave particles the
critical point is determined by
\begin{eqnarray}
&&\bar T_c=\frac{\bar U}2\biggl[\ln\frac{(\bar U-24)(\bar
U+3)}{(\bar
U-6)(\bar U+12)}\biggr]^{-1},~~{\rm (boson)} \label{be}\\
&&\bar T_c=\frac{\bar U}2\biggl[\ln\frac{(\bar U+12)(\bar
U+3)}{(\bar U-6)^2}\biggr]^{-1},~~{\rm (fermion)} \label{fe}
\end{eqnarray}
where $\bar T_c=T_c/zt$. There is no analytical solution when
$\alpha_M\geq3$, but the equations may be numerically solved. The
results of four types of the slave bosons have been presented in
Ref. \cite{dick}. We have solved the slave fermion equations for
$\alpha_M=3$. In Fig. 2, we plot the critical temperature as
a function of $U$. Below $T_c$, it is superfluid phase and above
$T_c$, it is normal liquid phase. $T_c=0$ is a triplet critical
point of the superfluid, normal liquid and Mott insulator phases.

We now make a comparison with the results in Fig. 2 from the two
approaches. The critical temperature for a given $U$ from the
slave boson approach is higher than that from the slave fermion
one. Two further analyzes can show that the slave fermion result
may be more appropriate. The bar located at $\bar T=1.18$ on the
$\bar T$-axis is the exact critical temperature for the ideal Bose
gas on the three-dimensional lattice in the long wave length limit
\cite{note2}. It is obvious that the critical temperature at $U=0$
from the slave fermion approach is better then that from the slave
boson approach.  As mentioned in \cite{dick}, there are two
unsatisfactory features for slave bosons: one is that the
zero-temperature critical interaction $U$ is moved to $U_c=6$ from
solving (\ref{cstde}) and (\ref{critical}) and is different from
zero-temperature result  $U_c\approx 5.83$ from given by (\ref{ml}).
Another is the existence of a local maximum in the position of the phase
boundary indicated by
the arrows in Fig. 2. We see that both features are improved by
the slave fermion approach. Although the three slave fermion
result still gives $U_c=6$ at $T=0$, the four fermion result has
moved $U_c$ to about 5.85, which is very close that the well-known
zero-temperature mean field result $\bar U_c\approx 5.83$. Note
that there are solutions of $U>6$ in eq. (\ref{fe}) where the
'critical temperatures' are nonzero, which are shown by filled
circles in Fig. 2. However, these solutions correspond to $n^1>1$
and $n^0<0$. Thus, these points are not physical. The similar
situation appears for four slave fermions but these points move to
$U>5.85$ (shown by the empty circles in Fig. 2). To verify these
'critical temperatures' are not physical, we depict $n^1$ at the
critical point as a function of $U$ in Fig. 3. We see that $n^1>1$
after $U>5.85$. All of these evidences show the critical point
indeed moves to around $U_c\sim 5.85$ at $T=0$, which is in good
agreement with the zero temperature mean field result, $\bar
U_c=5.83$.

The second feature in the slave boson result is the local maximum
in the $U$-$T_c$ curves (see the arrows in Fig. 2). As we see
soon, this maximum comes from the finite $\alpha_M$ approximation
getting worse as $U\to 0$. This feature is greatly improved in the
slave fermion $U$-$T_c$ curves. The positions of the maximum of
the critical temperature for slave fermions are in about $\bar
U\sim 1.2$ for three fermions and near $\bar U=0.2$ for four
fermions, which are much closer to zero than 2.15 and 1.8 in the
slave boson curves. Furthermore, $\Delta T=\bar T_c^{\rm max}-\bar
T_c(\bar U=0)\sim 0.25$ and $\Delta T/T_c^{\rm max}\sim 15\%$ for
slave bosons while $\Delta T\sim 0.02$ and $\Delta T/T_c^{\rm
max}\sim 2\%$ for slave fermions. The local maximum appears
because the cut-off $\alpha\leq 3$ is not appropriate as $U\to 0$.

Both these two quantitative improvements from the slave fermion
approach over the slave boson approximation come from the
exclusion of the same types of the slave fermions, due to the
Pauli principle. From (\ref{be}) and (\ref{fe}), we can calculate
the derivative $\frac{dT_c}{dU}$. It is seen that the boson
statistics of the slave boson sharpens the slope the $T_c$-$U$
curve in small $U$.

We next investigate if the finite $\alpha_M$ approximation is good
or not. For this purpose, we plot $n^\alpha$ ($\alpha_M=3$) versus
$U$ at the critical temperature (Fig. 4). It is seen that the
occupancy of the $0$th, 1st and 2nd types of the slave fermion is
of the order $10^{-1}$ from $U=1$ to 4. However, $n^3$ decreases
quickly as $U$ increases. For $U=1$, $n^3\approx 0.05$ but
$\approx 0.009,5\times 10^{-4}$ and $ 2\times 10^{-5}$ for $U=2,3$
and 4, respectively. This means that for a large enough $U$, the
$\alpha_M=3$ cut-off is a good approximation. In the regime of
small $U$, a larger $\alpha_M (>3)$ is required if we would like
to have a quantitatively reliable result. For $\alpha_M=3$, we see
that, from Fig. 2, $\bar T_c(U=0,\alpha_M=3)\approx 0.98$. It may
be expected that as $\alpha_M$ increases, $\bar T_c(U=0)$ should
be close to, e.g, 1.18 in three dimensions, the ideal Bose gas
critical temperature. The approximation getting worse for small
$U$ means the contribution from large $\alpha(>3)$ can not be
neglected.  The maximum of the critical temperature in Fig. 2
comes from neglecting these degrees of freedom corresponding to
large $\alpha$. Taking a larger $\alpha_M$, it may be anticipated
that the maximum of $\bar T_c$ may disappear as $\bar T_c(U=0)$
tends to 1.18. To reveal the quantitative behavior of the system
for small $U$ more precisely, we have to work at a larger
$\alpha_M$. The numerical work is still in progress
which will be present elsewhere.\\

\begin{figure}
\begin{center}
\includegraphics[width=8cm]{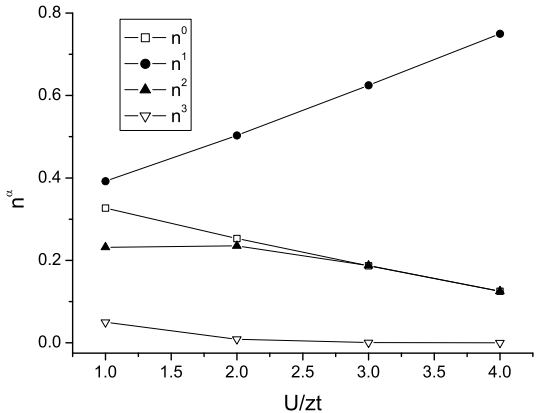}
\end{center}
 \caption{\label{fig:Fig. 4} The slave fermion occupant number $n^\alpha$
 versus $\bar U$ at the critical temperature ($\alpha_M=3$).}

\end{figure}

\begin{figure}
\begin{center}
\includegraphics[width=8cm]{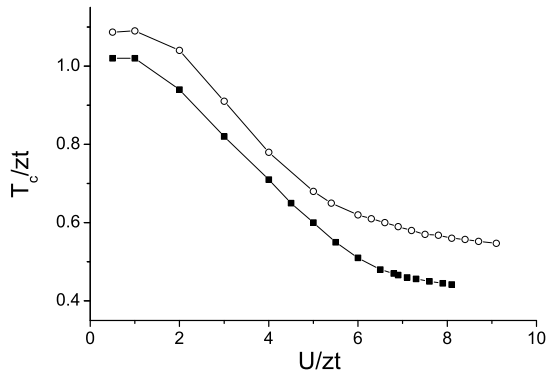}
\end{center}
 \caption{\label{fig:Fig. 5} The phase diagram for the
 incommensurate fillings. The filled squares are for $n=0.9$
 and the empty circles are for $n=0.75$. }
\end{figure}

\subsection{Incommensurate State}

It was also known that in the incommensurate state, there is no
Mott insulator phase for any value of $U$. To confirm this point,
we calculate two incommensurate fillings with $n=0.9$ and 0.75. As
Fig. 5 shown, only a normal-superfluid transition in the phase
diagram is found and there is no Mott insulator phase.

\section{conclusion}

In summary, we have made a comparative study of the finite
temperature phase diagrams with the slave boson and the slave
fermion mean field theory for the Bose Hubbard model. We found
that both slave particle mean field theories are qualitatively the
same but the slave fermion approach is quantitatively more
accurate. Many other results obtained in Ref. \cite{dick} by
the slave boson approach are valuable and may be improved by the
slave fermion approach by replacing the bosonic occupation
number with the the fermionic occupation number. This approach can
be generalized to the two-component Bose Hubbard model with $
b^\dagger_{i,\uparrow}
=\sum_{\alpha_\uparrow,\alpha_\downarrow}\sqrt{\alpha_\uparrow+1}c^\dagger_{\alpha_\uparrow+1,
\alpha_\downarrow}c_{\alpha_\uparrow, \alpha_\downarrow}$ and $
b^\dagger_{i,\downarrow}
=\sum_{\alpha_\uparrow,\alpha_\downarrow}\sqrt{\alpha_\downarrow+1
}c^\dagger_{\alpha_\uparrow,
\alpha_\downarrow+1}c_{\alpha_\uparrow, \alpha_\downarrow}$, while
for Bose-Fermi mixture Hubbard model, a slave boson-composite
fermion mixture approach, with
$b_i^\dagger=\sum_\alpha\sqrt{\alpha+1}
(c^\dagger_{i,\alpha+1}c_{i\alpha}
+b^\dagger_{i,\alpha+1}b_{i\alpha})$ and $f_i^\dagger=\sum_\alpha
c^\dagger_{i,\alpha}b_{i,\alpha}$ or $=\sum_\alpha
b^\dagger_{i,\alpha}c_{i,\alpha}$, will be required. Here
$c^\dagger_{i,\alpha}$ may be thought as a composite fermion
\cite{cf}. These will be worked out elsewhere.

This work was supported in part by Chinese National Natural
Sciences Foundation and NSF of the US.


\begin{references}
\bibitem{note1} To eliminate the multi-occupied configuration
beyond the mean field state, one has to deal with the fluctuation
corresponding to $\delta\lambda_i=\lambda_i-\lambda$. However, it
is very difficult to solve this problem.
\bibitem{note2} In the long wave length limit, the critical
temperature for the three-dimensional ideal Bose gas is given by
$\bar T^{ideal}_c=\frac{2\pi}{6}\biggl(\frac{\rho
a^3}{g_{3/2}(0)}\biggr)^{2/3}$ with
$g_i(x)=\sum_{m=1}^{\infty}\frac{e^{mx}}{m^i}$. For the present
case, $\rho a^3=n=1$ and then $\bar
T^{ideal}_c=2.61^{-2/3}\times\frac{4\pi}6\approx 1.18$.
\end{references}
\end{document}